\documentclass[preprint,12pt]{elsarticle}
\usepackage{amsmath}\usepackage{color}
\newtheorem{thm}{Theorem}

\begin{document}
\begin{frontmatter}
\title{A four-component Camassa-Holm type hierarchy}

\author{Nianhua Li$^1$, Q. P. Liu$^1$, Z. Popowicz$^2$}


\address{$^1$Department of Mathematics\\
China University  of Mining and Technology\\
Beijing 100083, P R China}

\address{$^2$Institute of Theoretical Physics\\ University of Wroc{\l}aw
\\ pl. M. Borna 9, 50-205\\ Wroc{\l}aw, Poland, ziemek@ift.uni.wroc.pl}

\begin{abstract}
We consider a 3$\times$3 spectral problem which generates four-component CH type systems. The bi-Hamiltonian structure and infinitely many conserved quantities are constructed for the associated hierarchy. Some possible reductions are also studied.
\bigskip

\noindent
Mathematical Subject Classification: 37K10, 37K05
\end{abstract}
\begin{keyword} Bi-Hamiltonian structure, Conserved quantity, Lax pair, Integrability.
\end{keyword}

\end{frontmatter}
\section{Introduction}
The Camassa-Holm (CH) equation
\begin{equation}\label{ch}
u_t-u_{xxt}+3uu_x=2u_xu_{xx}+uu_{xxx},
\end{equation}
has been a subject of steadily growing literature since it was derived from the incompressible Euler equation to model long waves in shallow water by Camassa and Holm in 1993 \cite{Holm}. It is a completely integrable system, which possesses Lax representation and is a bi-Hamiltonian system and  especially admits peakon solutions \cite{Holm}\cite{Hyman}.

CH equation is a system with quadratic nonlinearity, and another such system was proposed in 1999 by Degasperis and Procesi \cite{Degasperis}, which reads as
\begin{equation}\label{dpe}
m_{t}+um_{x}+3u_{x}m=0,\quad m=u-u_{xx},
\end{equation}
now known as called DP equation whose peakon solutions were studied in \cite{gas}\cite{szm}.   The following third order spectral problem for the DP equation \eqref{dpe} was found by Degasperis, Holm and Hone \cite{gas}
\begin{equation}
 \psi_{xxx} = \psi_{x} -\lambda \psi
\end{equation}
or equivalently it may be rewritten in the matrix form as
\begin{equation}
\varphi_{x}=\left(
              \begin{array}{ccc}
                0 & 0 & 1 \\
                -\lambda m & 0 & 0 \\
                1 & 1 & 0 \\
              \end{array}
            \right)\varphi.
\end{equation}

Olver and Rosenau suggested so-called tri-Hamiltoninan duality approach to construct the CH type equations and many examples were worked out  \cite{olver} (see also \cite{for}\cite{fokas}\cite{fuch}). In particular, they  obtained  a CH type equation with cubic nonlinearity
\begin{equation}\label{Olverek}
 m_t + [m(u^2 - u_x^2)]_x=0, ~~~~~ m=u-u_{xx}
\end{equation}
which, sometimes is referred as Qiao equation, was studied by Qiao \cite{Qiao}.

Recently, Song, Qu and Qiao  \cite{Qu} proposed a two-component generalization of (\ref{Olverek})
\begin{eqnarray}\label{song}
\left\{\begin{array}{rl}
m_{t} -[ m(u_xv_x -uv + uv_x - u_xv)]_x=0,& m=u-u_{xx},\\[8pt]
 n_{t} -[n(u_xv_x - uv + uv_x - u_xv)]_x=0,& n=v-v_{xx}\end{array} \right.
\end{eqnarray}
based on the following spectral problem
\begin{equation}\label{song1}
\varphi_{x}=\left(
              \begin{array}{cc}
                \frac{1}{2} & \lambda m \\
               \lambda n & -\frac{1}{2}  \\
              \end{array}
            \right)\varphi.
\end{equation}
This system \eqref{song1} is shown to be bi-Hamiltonian and is associated with WKI equation \cite{Tian}.

Moreover  Xia,Qiao and Zhou \cite{xia} generalized previous system to the 
integrable two-component CH type equation
\begin{eqnarray}\nonumber 
 m_t &=& F + F_x - \frac{1}{2}m(uv - u_xv_x + uv_x − u_xv),\nonumber \\
n_t &=& -G + G_x + \frac{1}{2}n (uv - u_xv_x + uv_x - u_xv) ,\\ \nonumber 
 m &=& u - u_{xx}, ~~~~~~~ n = v - v_{xx} \nonumber 
\end{eqnarray}
where F and G are two arbitrary functions.

Working on symmetry classification of nonlocal partial differential equations,  Novikov \cite{Novikov} found another  CH type equation with cubic nonlinearity. It  reads as
\begin{equation}\label{novikov}
m_{t}+u^{2}m_{x}+3uu_{x}m=0,\quad m=u-u_{xx}.
\end{equation}
Subsequently, Hone and Wang proposed a Lax representation for (\ref{novikov}) and showed that it is associated to a negative flow in Sawada-Kotera hierarchy \cite{hwang}. Furthermore, infinitely many conserved quantities and bi-Hamiltonian structure are also presented \cite{Hone}.

A two-component generalization  of  the  Novikov equation was constructed by Geng and Xue \cite{Geng}.
\begin{eqnarray}\label{nu}
\left\{\begin{array}{rl} m_{t}+3u_{x}vm+uvm_{x} =0,& m=u-u_{xx}\\[8pt]
n_{t}+3v_{x}un+uvn_{x} =0,&   n=v-v_{xx}\end{array}\right.,
\end{eqnarray}
with a spectral problem
\begin{equation}\label{}
\varphi_{x}=\left(
              \begin{array}{ccc}
                0 & \lambda m & 1 \\
                0 & 0 & \lambda n \\
                1 & 0 & 0 \\
              \end{array}
            \right)\varphi,
\end{equation}
which reduces  to the spectral Novikov's system as $m=n$. They also calculated the $N$-peakons and conserved quantities and found a Hamiltonian structure.

In this paper we will discuss the properties of the equations which follow from the
following generalized spectral problem
\begin{equation}\label{hk}
 \Phi_{x}=U\Phi, \quad U=\left(
    \begin{array}{ccc}
      0 & \lambda m_{1}& 1 \\
       \lambda n_{1}& 0 & \lambda m_{2}\\
      1 &  \lambda n_{2}& 0 \\
    \end{array}
  \right).
\end{equation}
This spectral problem was proposed by one of us (ZP) recently \cite{popowicz} and interestingly the resulted nonlinear systems can involve an arbitrary function. This freedom allows us to recover many known CH type equations from reductions.
As we will show this spectral problem takes the spectral problem considered by Geng and Xue for three-component system \cite{Geng2},  one and two-component Novikov's equation \cite{Hone}\cite{Geng},
and one or two component Song-Qu-Qiao equation \cite{Qu} as special cases.  In this sense, almost all known $3\times 3$ spectral problems for the CH type equations are contained in this case,  so it is interesting to study this spectral problem.

The  first negative flow corresponding to the spectral problem (\ref{hk}) is
\begin{eqnarray}\label{novikov2}\label{eqq}
\left\{\begin{array}{r}
  m_{1t}+n_{2}g_{1}g_{2}+m_{1}(f_{2}g_{2}+2f_{1}g_{1})=0,\\[5pt]
m_{2t}-n_{1}g_{1}g_{2}-m_{2}(f_{1}g_{1}+2f_{2}g_{2})=0,\\[5pt]
n_{1t}-m_{2}f_{1}f_{2}-n_{1}(f_{2}g_{2}+2f_{1}g_{1})=0,\\[5pt]
n_{2t}+m_{1}f_{1}f_{2}+n_{2}(f_{1}g_{1}+2f_{2}g_{2})=0,   \\[5pt]
m_{i}=u_{i}-u_{ixx},n_{i}=v_{i}-v_{ixx}, i=1,2, \end{array}\right.
\end{eqnarray}
where
\begin{eqnarray}
& f_{1}=u_{2}-v_{1x},\quad f_{2}=u_{1}+v_{2x},\nonumber\\
& g_{1}=v_{2}+u_{1x}, \quad g_{2}=v_{1}-u_{2x}.\nonumber
\end{eqnarray}
which will be shown to be a bi-Hamiltonian system.

The paper is organized as follows. In the section 2, we will construct bi-Hamiltonian operators related to the spectral problem \eqref{hk} and present a bi-Hamilotonian representation for the system \eqref{novikov2}. In the section 3, we construct infinitely many conserved quantities for the integrable hierarchy of \eqref{hk}. In the section 4, we consider the special  reductions of our  spectral problem.
The last section contains  concluding remarks.

\section{Bi-Hamiltonian structures in general}

Let us consider the following Lax pair
\begin{equation}\label{11}
 \Phi_{x}=U\Phi, \quad \Phi_{t}=V\Phi,
\end{equation}
where  $U$ is defined by (\ref{hk}) and $ V=\left(V_{ij}\right)_{3\times 3}$ at the moment is an arbitrary matrix.
The compatibility condition of (\ref{11}) or the zero-curvature representation
\begin{equation}\label{lv}
U_{t}-V_{x}+[U,V]=0,
\end{equation}
is equivalent to
\begin{equation}\label{zero}
\left\{\begin{array}{c}\lambda m_{1t}=V_{12x}-V_{32}+\lambda (m_{1}V_{11}+n_{2}V_{13}-m_{1}V_{22}),\\
\lambda m_{2t}=V_{23x}+V_{21}+\lambda (m_{2}V_{22}-m_{2}V_{33}-n_{1}V_{13}),\\
\lambda n_{1t}=V_{21x}+V_{23}+\lambda( n_{1}V_{22}-m_{2}V_{31}-n_{1}V_{11}),\\
\lambda n_{2t}=V_{32x}-V_{12}+\lambda (n_{2}V_{33}+m_{1}V_{31}-n_{2}V_{22}),\end{array}\right.
\end{equation}
with
\begin{eqnarray}
V_{11}&=&V_{31x}+V_{33}-\lambda n_{2}V_{21}+\lambda n_{1}V_{32},\nonumber\\
V_{13}&=&V_{33x}+V_{31}+\lambda m_{2}V_{32}-\lambda n_{2}V_{23},\nonumber\\
V_{22x}&=&\lambda(n_{1}V_{12}+m_{2}V_{32}-m_{1}V_{21}-n_{2}V_{23}),\label{vvv}\\
2V_{31x}+V_{33xx}&=&\lambda ((\partial n_{2}+m_{1})V_{23}-(\partial m_{2}+n_{1})V_{32}-m_{2}V_{12}+n_{2}V_{21}),\nonumber\\
2V_{33x}+V_{31xx}&=&\lambda ((\partial n_{2}+m_{1})V_{21}-(\partial n_{1} +m_{2})V_{32}-n_{1}V_{12}+n_{2}V_{23}).\nonumber
\end{eqnarray}
Taking account of (\ref{vvv}) and through a tedious calculation, the system (\ref{zero}) yields
\begin{equation}\label{master}
\left(\begin{array}{c}
m_{1}\\m_{2}\\n_{1}\\n_{2}
\end{array}\right)_t=
(\lambda^{-1}{\cal K}+\lambda {\cal L})
\left(\begin{array}{c}
V_{21}\\ V_{32}\\V_{12}\\V_{23}
\end{array}\right),
\end{equation}
with
\begin{eqnarray}\label{jj}
{\cal K}=\left(
           \begin{array}{cccc}
             0 & -1 & \partial & 0 \\
             1 & 0 & 0 & \partial \\
             \partial & 0 & 0 & 1 \\
             0 & \partial & -1 & 0 \\
           \end{array}
         \right)
,\quad {\cal L}={\cal J}+{\cal F},
\end{eqnarray}
and
\begin{eqnarray}
&&{\cal J}=\left(
                    \begin{array}{cccc}
                      2m_{1}\partial^{-1}m_{1}  & -m_{1}\partial^{-1}m_{2} & {\cal J}_{13} & {\cal J}_{14}\\
                       -m_{2}\partial^{-1}m_{1} &  2m_{2}\partial^{-1}m_{2} & {\cal J}_{23} &  {\cal J}_{24} \\
                       -{\cal J}_{13}^{\ast} &   -{\cal J}_{23}^{\ast} &  2n_{1}\partial^{-1}n_{1} &  -n_{1}\partial^{-1}n_{2} \\
                      -{\cal J}_{14}^{\ast} &  -{\cal J}_{24}^{\ast} &  -n_{2}\partial^{-1}n_{1} &  2n_{2}\partial^{-1}n_{2} \\
                    \end{array}
                  \right),\nonumber\\[8pt]
&&{\cal F}=(2P+S\partial)(\partial^{3}-4\partial)^{-1}P^{\mbox{\scriptsize{T}}}-
(2S+P\partial)(\partial^{3}-4\partial)^{-1}S^{\mbox{\scriptsize{T}}},\nonumber
\end{eqnarray}
where
\begin{eqnarray}
&&{\cal J}_{13}=-2m_{1}\partial^{-1}n_{1}-n_{2}\partial^{-1}m_{2},\; {\cal J}_{14}=m_{1}\partial^{-1}n_{2}+n_{2}\partial^{-1}m_{1},\nonumber\\
&&{\cal J}_{23}=m_{2}\partial^{-1}n_{1}+n_{1}\partial^{-1}m_{2},
\;\; \quad {\cal J}_{24}=-2m_{2}\partial^{-1}n_{2}-n_{1}\partial^{-1}m_{1},\nonumber\\
&& P=(m_1,m_2,-n_1,n_2)^{\mbox{\scriptsize{T}}}, \;\; \quad\quad\quad\; S=(-n_2,n_1,-m_2,m_1)^{\mbox{\scriptsize{T}}}.\nonumber
\end{eqnarray}

By specifying the matrix $V$ properly, we can obtain the hierarchy of equations associated with the Lax pair \eqref{11}. For example, if we assume
\begin{eqnarray}\label{lax}
&&V=\left(
    \begin{array}{ccc}
      -f_{1}g_{1} & \frac{g_{1}}{\lambda} & -g_{1}g_{2} \\[5pt]
      \frac{f_{1}}{\lambda}& -\frac{1}{\lambda^{2}}+f_{1}g_{1}+f_{2}g_{2} & \frac{g_{2}}{\lambda} \\[5pt]
      -f_{1}f_{2} & \frac{f_{2}}{\lambda} & -f_{2}g_{2} \\
    \end{array}
  \right), \nonumber
 \end{eqnarray}
then the resulted system,  the first negative flow of the hierarchy, is exactly the equation (\ref{novikov2})  .
On the other hand  taking the $V$ matrix in time part of the Lax pair as
\begin{eqnarray}
& \tilde{V}=-\lambda \left(
    \begin{array}{ccc}
      0 &  m_{1}\Gamma & 0 \\
       n_{1}\Gamma & 0 &  m_{2}\Gamma \\
      0 &  n_{2}\Gamma & 0 \\
    \end{array}
  \right).
\end{eqnarray}
we obtain
\begin{eqnarray}\label{fyrst}
\left\{ \begin{array}{rr}m_{1t}+(\Gamma m_{1})_{x}-n_{2}\Gamma=0,&
m_{2t}+(\Gamma m_{2})_{x}+n_{1}\Gamma=0,\\[5pt]
n_{1t}+(\Gamma n_{1})_{x}+m_{2}\Gamma=0,& n_{2t}+(\Gamma n_{2})_{x}-m_{1}\Gamma=0,
\end{array}\right.\end{eqnarray}
where $\Gamma$ is an arbitrary function.

Now the combinations of  $V$ and $\tilde{V}$  leads us to the following system of equations
\begin{eqnarray}\label{newequ}
\left\{\begin{array}{r} m_{1t}+(\Gamma m_{1})_{x}+n_{2}(g_{1}g_{2}-\Gamma)+m_{1}(f_{2}g_{2}+2f_{1}g_{1})=0,\\[5pt]
m_{2t}+(\Gamma m_{2})_{x}-n_{1}(g_{1}g_{2}-\Gamma)-m_{2}(f_{1}g_{1}+2f_{2}g_{2})=0, \\[5pt]
 n_{1t}+(\Gamma n_{1})_{x}-m_{2}(f_{1}f_{2}-\Gamma)-n_{1}(f_{2}g_{2}+2f_{1}g_{1})=0,\\[5pt]
 n_{2t}+(\Gamma n_{2})_{x}+m_{1}(f_{1}f_{2}-\Gamma)+n_{2}(f_{1}g_{1}+2f_{2}g_{2})=0,\\[5pt]
 m_{i}~=~u_{i}-u_{ixx},~~n_{i}~=~v_{i}-v_{ixx},~~ i=1,2,\end{array}\right.
\end{eqnarray}
which has the Lax pair
\begin{equation}\label{}
\Phi_{x}=U\Phi,\quad \Phi_{t}=(V+\tilde{V})\Phi.\nonumber
\end{equation}


It is obvious that the operator ${\cal K}$ given by \eqref{jj} is a Hamiltonian operator. Indeed we have the following theorem
\begin{thm} The operators ${\cal K}$ and ${\cal L}$ defined  by (\ref{jj}) constitute a pair of compatible Hamiltonian operators. In particular, the four-component system (\ref{eqq}) is a bi-Hamiltonian system, namely it can be written as
\begin{equation}\label{}
  \left(
   \begin{array}{c}
     m_{1} \\
     m_{2} \\
     n_{1}\\
     n_{2}\end{array}
 \right)_{t}={\cal K}\left(
                       \begin{array}{c}
                          \frac{\delta H_{0}}{\delta m_{1}} \\[5pt]
                          \frac{\delta H_{0}}{\delta m_{2}} \\[5pt]
                          \frac{\delta H_{0}}{\delta n_{1}}\\[5pt]
                          \frac{\delta H_{0}}{\delta n_{2}} \\
                       \end{array}
                     \right)
 ={\cal L}\left(
            \begin{array}{c}
               \frac{\delta H_{1}}{\delta m_{1}} \\[5pt]
               \frac{\delta H_{1}}{\delta m_{2}} \\[5pt]
               \frac{\delta H_{1}}{\delta n_{1}}\\[5pt]
               \frac{\delta H_{1}}{\delta n_{2}}\\
            \end{array}
          \right)
\end{equation}
where
\begin{eqnarray}
H_{0}=\int{(f_{1}g_{1}+f_{2}g_{2})(m_{2}f_{2}+n_{1}g_{1})}dx, \quad
H_{1}=\int {(m_{2}f_{2}+n_{1}g_{1})} dx.
\end{eqnarray}
\end{thm}
\noindent
{\bf Proof}: It is easy to check that  ${\cal L}$ is a skew-symmetric operator. Thus, what we need to do is to verify the Jacobi identity  for ${\cal L}$ and the compatibility of  two operators ${\cal K}$ and ${\cal L}$. To this end, we follow Olver and use his multivector approach \cite{PJOlver}. Let us introduce
\begin{eqnarray}
 \Theta_{\cal K}=\frac{1}{2}\int {\theta\wedge{\cal K}\theta}dx,\quad \Theta_{\cal L}=\frac{1}{2}\int {\theta\wedge{\cal L}\theta}  dx
\end{eqnarray}
with $\theta=(\theta_{1},\theta_{2},\theta_{3},\theta_{4})$. A tedious but direct calculation shows that
\begin{eqnarray}
 \mbox{pr}\; v_{{\cal L}\theta}(\Theta_{\cal L})=0,\quad \mbox{pr}\; v_{{\cal K}\theta}(\Theta_{\cal L})+\mbox{pr}\; v_{{\cal L}\theta}(\Theta_{\cal K})=0
\end{eqnarray}
hold. The details of this calculation are  postponed to the Appendix.

\bigskip

Since we have a bi-Hamiltonian pair ${\cal K}$ and ${\cal L}$, we may formulate an integrable hierarchy of the corresponding nonlinear evolutions equations through  recursion operator.
While the  system  (\ref{fyrst}) possesses a Lax pair, we could not expect that it is a bi-Hamiltonian system for any $\Gamma$ and the constraint of $\Gamma$ will appear if it is requested to commute with the negative flow .

To understand the appearance of the arbitrary function occurred in the present case, we now calculate the Casimir functions of the Hamiltonian operator ${\cal L}$. Let
\begin{equation}\label{ker}
{\cal L}(A,B,C,D)^{\mbox{\scriptsize{T}}}=0,
\end{equation}
 and define
\begin{eqnarray}\label{line}
&&K_{1}=m_{1}A-n_{1}C,\quad\;\;  K_{2}=m_{2}B-n_{2}D,\\ &&K_{3}=m_{2}C-m_{1}D,\quad K_{4}=n_{2}A-n_{1}B.\label{ljj}
 \end{eqnarray}
The system \eqref{ker} consists of four equations of similar type. For example the one of them is
\begin{eqnarray*}
\lefteqn{m_{1}\left(\partial^{-1}(2K_{1}-K_{2})+(\partial^{3}-4\partial)^{-1}(2(K_{1}+K_{2})+(K_{3}+K_{4})_{x}\right)=}\\
&&{}\qquad \qquad\qquad n_{2}(\partial^{-1}K_{3}+\partial^{3}-4\partial)^{-1}(2(K_{3}+K_{4})+(K_{1}+K_{2})_{x})).
\end{eqnarray*}
Solving these equations we found
\begin{eqnarray*}
K_{1}&=&(m_{2}n_{2}\Lambda)_{x}+(n_{1}n_{2}-m_{1}m_{2})\Lambda, \\ K_{2}&=&-(m_{1}n_{1}\Lambda)_{x}+(n_{1}n_{2}-m_{1}m_{2})\Lambda, \\
K_{3}&=&(m_{1}m_{2}\Lambda)_{x}+(m_{1}n_{1}-m_{2}n_{2})\Lambda, \\ K_{4}&=&-(n_{1}n_{2}\Lambda)_{x}+(m_{1}n_{1}-m_{2}n_{2})\Lambda,
\end{eqnarray*}
where $\Lambda=\frac{k}{m_{1}n_{1}+m_{2}n_{2}}$  and $k$ is an arbitrary number. Substituting above expressions for $K_i$ into \eqref{line}-\eqref{ljj} and solving the resulted linear equations leads to
\begin{eqnarray}
&&A=-n_{1}\Gamma+\frac{n_{1}}{m_{1}m_{2}}K_{3}+\frac{1}{m_{1}}K_{1},\nonumber\\
&&B=-n_{2}\Gamma+\frac{1}{m_{2}}K_{2},\quad
C=-m_{1}\Gamma+\frac{1}{m_{2}}K_{3}, \quad D=-m_{2}\Gamma,\nonumber
\end{eqnarray}
where $\Gamma$ is an arbitrary function. This implies that ${\cal L}$ is a degenerate Hamiltonian operator. For the special case $k=0$ and $\Gamma = m_1 n_1 + m_2 n_2$,  the Casimir is $H_c = -\frac{1}{2}\int{\Gamma^2 dx}  $ and hence the system (\ref{fyrst}) in this case is a Hamiltonian system
\begin{equation*}
  \left(
   \begin{array}{c}
     m_{1} \\
     m_{2} \\
     n_{1}\\
     n_{2}\end{array}
 \right)_{t}={\cal K}\left(
                       \begin{array}{c}
                          \frac{\delta H_{c}}{\delta m_{1}} \\[5pt]
                          \frac{\delta H_{c}}{\delta m_{2}} \\[5pt]
                          \frac{\delta H_{c}}{\delta n_{1}}\\[5pt]
                          \frac{\delta H_{c}}{\delta n_{2}} \\
                       \end{array}
                     \right).
 \end{equation*}

\section{Conserved quantities}
An integrable system normally possesses infinity number of concerned quantities and such property has been taken as one of the defining properties for integrability. In this section, we show that  infinitely many conserved quantities can be  constructed for the nonlinear evolution equations related with the spectral problem (\ref{hk}). Indeed, we may derive two sequences of conserved quantities  utilizing   the projective coordinates in  the spectral problem. We can introduce these  coordinates in three different manners as
\begin{equation} \nonumber
 \textbf{ I.)}~~  a=\frac{\varphi_{1}}{\varphi_{2}}, ~\quad b=\frac{\varphi_{3}}{\varphi_{2}},\hspace{1cm}
  \textbf{II.)}~~  \sigma=\frac{\varphi_{2}}{\varphi_{1}},~\tau=\frac{\varphi_{3}}{\varphi_{1}},\hspace{1cm}
  \textbf{III.)}~~  \alpha=\frac{\varphi_{1}}{\varphi_{3}}, ~\beta=\frac{\varphi_{2}}{\varphi_{3}}
\end{equation}

{\bf Case 1:}
In these coordinates we obtain that
\begin{equation}\label{cc}
\rho=(\ln\varphi_{2})_{x}=\lambda n_{1}a+\lambda m_{2}b,
\end{equation}
is conserved quantity with $a, b$ satisfy
\begin{equation}\label{clll}
a_{x}=\lambda m_{1}+b-a\rho,\quad b_{x}=a+\lambda n_{2}-b\rho.
\end{equation}
Substituting the Laurent series expansions in $\lambda$ of  $a$ and $b$ into \eqref{clll}
\begin{equation}\label{}
a=\sum_{i\geq0}a_{i}\lambda^{i}, \quad b=\sum_{j\geq0}b_{j}\lambda^{j},\nonumber
\end{equation}
then we find
\begin{eqnarray}
&a_{0}=0, \; a_{1}=-v_{2}-u_{1x}=-g_{1}, \; a_{2}=0,\nonumber\\
&b_{0}=0, \; b_{1}=-u_{1}-v_{2x}=-f_{2}, \; b_{2}=0,\nonumber
\end{eqnarray}
and
\begin{eqnarray}
&&a_{k,x}=b_{k}-\sum_{i+j=k-1}(n_{1}a_{i}a_{j}+m_{2}a_{i}b_{j}),\nonumber\\
&&b_{k,x}=a_{k}-\sum_{i+j=k-1}(n_{1}a_{i}b_{j}+m_{2}b_{i}b_{j}),\; (k\geq3).\nonumber
\end{eqnarray}
With the aid of $a_{1},b_{1}$, we obtain a simple  conserved quantity
\begin{equation}\label{}
\rho_{1}=-\int (n_{1}g_{1}+m_{2}f_{2}) dx.
\end{equation}
Also, due to
\begin{eqnarray}
a_{3}-a_{3xx} &=& n_{1}f_{2}g_{1}+m_{2}f_{2}^{2}+(n_{1}g_{1}^{2}+m_{2}f_{2}g_{1})_{x}, \nonumber\\
b_{3}-b_{3xx} &=& n_{1}g_{1}^{2}+m_{2}f_{2}g_{1}+(n_{1}f_{2}g_{1}+m_{2}f_{2}^{2})_{x}, \nonumber
\end{eqnarray}
we obtain the next  conserved quantity by
\begin{eqnarray}
\rho_{3}&=&\int n_{1}a_{3}+m_{2}b_{3} dx=\int \big(v_{1}(a_{3}-a_{3xx})+u_{2}(b_{3}-b_{3xx})\big)dx \nonumber\\
&=&\int (n_{1}g_{1}+m_{2}f_{2})(f_{1}g_{1}+f_{2}g_{2})dx.
\end{eqnarray}

In addition, we may consider alternative expansions of $a, b$  in negative powers of $\lambda$, namely
\begin{equation*}\label{}
 a=\Sigma^{\infty}_{i\geq0}\tilde{a}_{i}\lambda^{-i},\quad  b=\Sigma^{\infty}_{j\geq0}\tilde{b}_{j}\lambda^{-j}.
\end{equation*}
As above, inserting these expansions into \eqref{clll} we may find   recursive relations for $\tilde{a}_{i},\tilde{b}_{j}$. The first two  conserved quantities are
\begin{eqnarray}
\rho_{0}&=&\int\sqrt{m_{1}n_{1}+m_{2}n_{2}}dx,\\
\rho_{-1}&=&\int\frac{2m_{1}m_{2}+2n_{1}n_{2}+m_{1}n_{1x}-m_{1x}n_{1}+m_{2x}n_{2}-m_{2}n_{2x}}{4(m_{1}n_{1}+m_{2}n_{2})}dx.\nonumber
\end{eqnarray}

{\bf Case 2:} The quantity $\bar{\rho}$ defined as
\begin{equation}\label{}
\bar{\rho}=(\ln\varphi_{1})_{x}=\lambda m_{1}\sigma+\tau,
\end{equation}
with $\sigma,\tau$ satisfy
\begin{equation}
\sigma_{x}=\lambda n_{1}+\lambda m_{2}\tau-\sigma \bar{\rho},\quad \tau_{x}=1+\lambda n_{2}\sigma-\tau \bar{\rho}.
\end{equation}
is conserved quantity.
Expanding $\sigma$ and $\tau$ in Laurent series of $\lambda$
then once again we may find the corresponding conserved quantities. For instance, in the case $k\geq0$, we get
\begin{equation}\label{}
 \bar{\rho_{2}}=\frac{1}{2}\int (m_{1}+n_{2})(f_{1}+g_{2})dx,\nonumber
\end{equation}
while in the case  $k\leq0$, we obtain
\begin{equation}\label{}
\bar{\rho}_{-1}=\int\frac{2m_{2}n_{2}^2+2m_{1}n_{1}n_{2}-m_{2x}m_{1}n_{2}+4m_{1x}m_{2}n_{2}-3n_{2x}m_{1}m_{2}+m_{1x}m_{1}n_{1}-n_{1x}m_{1}^2}{4m_{1}(m_{1}n_{1}+m_{2}n_{2})}dx.\nonumber
\end{equation}

{\bf Case 3:} For this case the conserved quantity is defined as

\begin{equation}\label{}
\hat{\rho}=(\ln\varphi_{3})_{x}=\alpha+\lambda n_{2}\beta,
\end{equation}
with $\alpha,\beta$ satisfy
\begin{equation}\label{alp}
\alpha_{x}=\lambda m_{1}\beta+1-\alpha \hat{\rho},\quad \beta_{x}=\lambda n_{1}\alpha+\lambda m_{2}-\beta\hat{\rho}.
\end{equation}

Expanding $\alpha$ and $\beta$ in Laurent series of $\lambda$
and substituting them into \eqref{alp}, we may obtain the conserved quantities and apart from those found in last two cases, we have
\begin{eqnarray}
\hat{\rho}_{-1}&=&\int\frac{2n_{1}n_{2}^2+2m_{1}m_{2}n_{2}-m_{2x}n_{2}^{2}+4n_{2x}m_{1}n_{1}-3m_{1x}n_{1}n_{2}+n_{2x}m_{2}n_{2}-n_{1x}m_{1}n_{2}}{4n_{2}(m_{1}n_{1}+m_{2}n_{2})}dx.\nonumber
\end{eqnarray}

Let us remark that these conserved quantities have been obtained from the $x$-part of the Lax pair representation only hence they are valid for the whole hierarchy. As we checked they are conserved for the system  (\ref{novikov2}) as well as for the (\ref{newequ}).

\section{Reductions}
We now consider the possible reductions of our four component spectral problem \eqref{11} and relate them to the spectral problems known in literatures.

\subsection{a three component reduction}

Assuming $m_1=u_1=0$, we have
\begin{equation}\label{33spectr}
\left(
  \begin{array}{c}
    \phi_1 \\
    \phi_2 \\
    \phi_3 \\
  \end{array}
\right)_{x}=\left(
              \begin{array}{ccc}
                0 & 0 & 1 \\
                \lambda n_1 & 0 & \lambda m_2 \\
                1 & \lambda n_2 & 0 \\
              \end{array}
            \right)\left(
  \begin{array}{c}
    \phi_1 \\
    \phi_2 \\
    \phi_3 \\
  \end{array}
\right),
\end{equation}
which, by identifying the variables as follows
\begin{equation}\label{vff}
n_{2}=u,\quad m_{2}=\frac{v}{u},\quad n_{1}=w+\left(\frac{v}{u}\right)_{x},
\end{equation}
may be reformulated as
\begin{equation*}
\left(
  \begin{array}{c}
   \varphi_1 \\
    \varphi_2 \\
    \varphi_3 \\
  \end{array}\right)_{x}
  =\left(
              \begin{array}{ccc}
                0 & 1 & 0 \\
                1+\lambda^2 v & 0 & u \\
                 \lambda^2 w & 0 & 0 \\
              \end{array}
            \right)\left(  \begin{array}{c}  \varphi_1 \\
    \varphi_2 \\     \varphi_3 \\   \end{array} \right).
\end{equation*}
 This spectral problem is nothing but the one proposed  by Geng and Xue \cite{Geng2} and the associated integrable flows are also bi-Hamiltonian \cite{Li}.  It is easy to see that \eqref{vff} is an invertible transformation, so to find the bi-Hamiltonian structures of the flows resulted from the spectral problem \eqref{33spectr} we may convert those for the Geng-Xue spectral problem into the present case. Direct calculations yield
\begin{eqnarray*}
&&{\cal L}_{\rm1}=\left(
                  \begin{array}{ccc}
                    -\frac{m_2}{n_2}\partial-\partial \frac{m_2}{n_2}& (\frac{m_2}{n_2}\partial+\partial \frac{m_2}{n_2})\partial+S & 0 \\
                    -\partial(\frac{m_2}{n_2}\partial+\partial \frac{m_2}{n_2})-S^{\ast} & \partial S+S^{\ast}\partial+\partial(\frac{m_2}{n_2}\partial+\partial \frac{m_2}{n_2})\partial & 1-\partial^2 \\
                    0 & \partial^2-1 & 0 \\
                  \end{array}
                \right),
\\
&&{\cal L}_{\rm2}=-\frac{1}{2}\left(
                              \begin{array}{c}
                                m_2\partial+2m_{2x} \\
                                m_2\partial^2+3n_1\partial+2n_{1x} \\
                                3n_2\partial+2n_{2x} \\
                              \end{array}
                            \right)(\partial^3-4\partial)^{-1}\left(
                              \begin{array}{c}
                                m_2\partial+2m_{2x} \\
                                m_2\partial^2+3n_1\partial+2n_{1x} \\
                                3n_2\partial+2n_{2x} \\
                              \end{array}
                            \right)^{\ast}\\
&&\hspace{0.9cm}+\frac{1}{2}\left(
   \begin{array}{ccc}
    3m_2\partial^{-1}m_2 & -m_{2}^{2}+3m_2\partial^{-1}n_1 & -3m_2\partial^{-1}n_2 \\
     m_{2}^{2}+3n_1\partial^{-1}m_2 & m_2\partial m_2+3n_1\partial^{-1}n_1 & -m_2n_2-3n_1\partial^{-1}n_2\\
     -3n_2\partial^{-1}m_2 & m_2n_2-3n_2\partial^{-1}n_1 & 3n_2\partial^{-1}n_2 \\
   \end{array}
 \right),
\end{eqnarray*}
where $S=\frac{m_2}{n_2}(1-\partial^2)$. We remark that it is not clear how to obtain above Hamilitonian pair from \eqref{jj}.

\noindent
\subsection{a two-component reduction}

In this case, we assume
\[
n_1=m_2,\quad n_2=m_1
\]
or
\begin{equation*}\label{reduc2}
\left(
  \begin{array}{c}
    \phi_1 \\
    \phi_2 \\
    \phi_3 \\
  \end{array}
\right)_{x}=\left(
              \begin{array}{ccc}
                0 &  \lambda m_1 & 1 \\
                \lambda m_2 & 0 & \lambda m_2 \\
                1 & \lambda m_1 & 0 \\
              \end{array}
            \right)\left(
  \begin{array}{c}
    \phi_1 \\
    \phi_2 \\
    \phi_3 \\
  \end{array}
\right),
\end{equation*}
which yields
\begin{equation}\label{reduc2}
\begin{pmatrix} \phi_1+\phi_3\\
\phi_2\end{pmatrix}_x=\begin{pmatrix}1&2\lambda m_1\\ \lambda m_2& 0\end{pmatrix}\begin{pmatrix}\phi_1+\phi_3\\ \phi_2\end{pmatrix}.
\end{equation}

By the following change of variables
\[
\phi_1+\phi_3= e^{\frac{1}{2}x}\psi_1,\; \phi_2= e^{\frac{1}{2}x}\psi_2, \; 2m_1= m,\; m_2= n,
\]
\eqref{reduc2} gives
\[
\begin{pmatrix} \psi_1\\
\psi_2\end{pmatrix}_x=\begin{pmatrix}\frac{1}{2}&\lambda m\\ \lambda n& -\frac{1}{2}\end{pmatrix}\begin{pmatrix}\phi_1\\ \psi_2\end{pmatrix},
\]
a spectral problem considered by Song, Qu and Qiao \cite{Qu}. Therefore the bi-hamiltonian structure of Song-Qu-Qiao  system (see \cite{Tian}) is hidden in our bi-hamiltonian structure.

\section{Concluding remarks}

In this paper,  started from a general $3\times 3$ problem, we  considered   the related four-component CH type systems. We obtained the  bi-Hamiltonian structure and suggested the way to construction of infinitely many conserved quantities for the integrable equations. Different reductions were also considered.

As noticed above, the positive flows allow for an arbitrary function $\Gamma$ involved and such systems are interesting since different specifications of $\Gamma$ lead to different CH type equations. Although the flow equations with arbitrary $\Gamma$ do possess infinitely many conserved quantities, we do not expect they are (bi-) Hamiltonian systems in general case.  Also, we explained the appearance of this arbitrary function by studying the kernel of one of the Hamiltonian operators and it seems that further study of such systems is needed.

A remarkable property of CH type equations is that it possess peakon solutions. One may find that the first negative flow, which does not depend on $\Gamma$,  only possess stationary peakons.
The systems such as \eqref{newequ} may admit non-stationary peakon solutions. This and other related issues may be considered in further publication.

\appendix

\renewcommand\thesection{\appendixname~\Alph{section}}

\renewcommand\theequation{\Alph{section}\arabic{equation}}

\section{}

We first prove that  ${\cal L}$ is also a Hamiltonian operator. For this purpose, we first define
\[
\Theta_{\cal J}=\frac{1}{2}\int {\theta \wedge \cal{J}\theta} dx, \; \Theta_{\cal F}=\frac{1}{2}\int {\theta \wedge \cal{F}\theta} dx, \; {\cal A}=(\partial^{3}-4\partial)^{-1}.
\]
By direct calculation, we have
\begin{eqnarray}
&&\hspace{0cm}\Theta_{\cal J}=\int \left((n_{2}\theta_{1}-n_{1}\theta_{2})\wedge \partial^{-1}(m_{1}\theta_{4}-m_{2}\theta_{3})+(m_{1}\theta_{1}-n_{1}\theta_{3})\wedge\partial^{-1}(n_{2}\theta_{4}-m_{2}\theta_{2})\right.\nonumber\\
&&\hspace{1.2cm}+\left.(n_{2}\theta_{4}-m_{2}\theta_{2})\wedge\partial^{-1}(n_{2}\theta_{4}-m_{2}\theta_{2})+(m_{1}\theta_{1}-n_{1}\theta_{3})\wedge\partial^{-1}(m_{1}\theta_{1}-n_{1}\theta_{3})\right)dx,\nonumber
\end{eqnarray}
and
\begin{eqnarray}
&&\hspace{0cm}\Theta_{\cal F}=\frac{1}{2}\int \big((2Q-R\partial)\wedge {\cal{A}}Q+(Q\partial-2R)\wedge{\cal{A}}R\big)dx\nonumber\\
&&\hspace{0.7cm}=\int \big(Q\wedge {\cal{A}}Q-R\wedge{\cal{A}}R+Q\wedge\partial{\cal{A}}R\big) dx,\nonumber
\end{eqnarray}
where
\begin{align*}
Q=m_{1}\theta_{1}+m_{2}\theta_{2}-n_{1}\theta_{3}-n_{2}\theta_{4}, \;
R=n_{2}\theta_{1}-n_{1}\theta_{2}+m_{2}\theta_{3}-m_{1}\theta_{4}.
\end{align*}

Since
\begin{eqnarray}
&&\mbox{pr}\,v_{\cal L \theta}(\Theta_{\cal L})=\mbox{pr}\,v_{\cal L \theta}(\Theta_{\cal J})+\mbox{pr}\,v_{\cal L \theta}(\Theta_{\cal F}),
\end{eqnarray}
we calculate $\mbox{pr}\,v_{\cal L \theta}(\Theta_{\cal J})$ and $\mbox{pr}\; v_{\cal L \theta}(\Theta_{\cal F})$. Indeed, we have
\begin{eqnarray*}
-\mbox{pr}\,v_{\cal L \theta}(\Theta_{\cal J})&=&\int\left[(\partial^{-1}(n_{2}\theta_{1}-n_{1}\theta_{2})\wedge\partial^{-1}(m_{1}\theta_{4}-m_{2}\theta_{3})\wedge\partial^{-1}Q)_{x} \right. \nonumber\\
&&+(2m_{1}\theta_{1}+2n_{1}\theta_{3}-n_{2}\theta_{4}-m_{2}\theta_{2})\wedge({\cal{A}}R_{x}+2{\cal{A}}Q)\wedge\partial^{-1}(m_{1}\theta_{1}-n_{1}\theta_{3})\nonumber\\
&&+(2m_{2}\theta_{3}-2n_{2}\theta_{1}+m_{1}\theta_{4}-n_{1}\theta_{2})\wedge({\cal{A}}Q_{x}+2{\cal{A}}R)\wedge\partial^{-1}(m_{1}\theta_{1}-n_{1}\theta_{3})\nonumber\\
&&+(2m_{1}\theta_{4}-2n_{1}\theta_{2}+m_{2}\theta_{3}-n_{2}\theta_{1})\wedge({\cal{A}}Q_{x}+2{\cal{A}}R)\wedge\partial^{-1}(n_{2}\theta_{4}-m_{2}\theta_{2})\nonumber\\
&&+(m_{1}\theta_{1}+n_{1}\theta_{3}-2m_{2}\theta_{2}-2n_{2}\theta_{4})\wedge({\cal{A}}R_{x}+2{\cal{A}}Q)\wedge\partial^{-1}(n_{2}\theta_{4}-m_{2}\theta_{2})\nonumber\\
&&+(m_{1}\theta_{4}-m_{2}\theta_{3})\wedge({\cal{A}}R_{x}+2{\cal{A}}Q)\wedge\partial^{-1}(n_{2}\theta_{1}-n_{1}\theta_{2})\nonumber\\
&&-(n_{1}\theta_{3}+n_{2}\theta_{4})\wedge({\cal{A}}Q_{x}+2{\cal{A}}R)\wedge\partial^{-1}(n_{2}\theta_{1}-n_{1}\theta_{2})\nonumber\\
&&+(m_{1}\theta_{1}+m_{2}\theta_{2})\wedge({\cal{A}}Q_{x}+2{\cal{A}}R)\wedge\partial^{-1}(m_{1}\theta_{4}-m_{2}\theta_{3})\nonumber\\
&&-\left.(n_{2}\theta_{1}-n_{1}\theta_{2})\wedge({\cal{A}}R_{x}+2{\cal{A}}Q)\wedge\partial^{-1}(m_{1}\theta_{4}-m_{2}\theta_{3})\right]dx.\nonumber
\end{eqnarray*}

Next we consider $\mbox{pr}\; v_{\cal L \theta}(\Theta_{\cal F})$. For simplicity, we denote
\[ \Upsilon=2m_{1}\theta_{1}-m_{2}\theta_{2}-2n_{1}\theta_{3}+n_{2}\theta_{4},\;\Omega=m_{1}\theta_{1}-2m_{2}\theta_{2}-n_{1}\theta_{3}+2n_{2}\theta_{4},\]
then a direct calculation shows that $\mbox{pr}\,v_{\cal L \theta}(\Theta_{\cal F})$ can be expressed as
\begin{eqnarray*}
-\mbox{pr}\,v_{\cal L \theta}(\Theta_{\cal F})&=&\int\left[(m_{1}\theta_{1}+n_{1}\theta_{3})\wedge\partial^{-1}\Upsilon\wedge(2{\cal{A}}Q+{\cal{A}}R_{x})\right.\nonumber\\
&&+2(n_{2}\theta_{1}-n_{1}\theta_{2}-m_{2}\theta_{3}+m_{1}\theta_{4})\wedge({\cal A}R_{x}+2{\cal A}Q)\wedge({\cal A}Q_{x}+2{\cal A}R)  \nonumber\\
&&-(m_{2}\theta_{2}+n_{2}\theta_{4})\wedge\partial^{-1}\Omega\wedge(2{\cal{A}}Q+{\cal{A}}R_{x}) \nonumber\\
&&+(n_{2}\theta_{1}-n_{1}\theta_{2})\wedge\partial^{-1}(m_{1}\theta_{4}-m_{2}\theta_{3})\wedge(2{\cal{A}}Q+{\cal{A}}R_{x})\nonumber\\
&&-(m_{1}\theta_{4}-m_{2}\theta_{3})\wedge\partial^{-1}(n_{2}\theta_{1}-n_{1}\theta_{2})\wedge(2{\cal{A}}Q+{\cal{A}}R_{x})\nonumber\\
&&+(m_{1}\theta_{4}-n_{1}\theta_{2})\wedge\partial^{-1}\Upsilon\wedge(2{\cal{A}}R+{\cal{A}}Q_{x})\nonumber\\
&&-(n_{2}\theta_{1}-m_{2}\theta_{3})\wedge\partial^{-1}\Omega\wedge(2{\cal{A}}R+{\cal{A}}Q_{x})\nonumber\\
&&+(n_{2}\theta_{4}+n_{1}\theta_{3})\wedge\partial^{-1}(m_{1}\theta_{4}-m_{2}\theta_{3})\wedge(2{\cal{A}}R+{\cal{A}}Q_{x})\nonumber\\
&&\left.-(m_{2}\theta_{2}+m_{1}\theta_{1})\wedge\partial^{-1}(n_{2}\theta_{1}-n_{1}\theta_{2})\wedge(2{\cal{A}}R+{\cal{A}}Q_{x})\right]dx.\nonumber
\end{eqnarray*}
Letting $f=n_{2}\theta_{1}-n_{1}\theta_{2},g=m_{1}\theta_{4}-m_{2}\theta_{3}$ and substituting above expansions into \eqref{?}
lead to
\begin{eqnarray*}
-\mbox{pr}\,v_{\cal L \theta}(\Theta_{\cal L})&=&\int2(n_{2}\theta_{1}-n_{1}\theta_{2}-m_{2}\theta_{3}+m_{1}\theta_{4})\wedge({\cal A}R_{x}+2{\cal A}Q)\wedge({\cal{A}}Q_{x}+2{\cal A}R)\nonumber\\
&&+2((n_{2}\theta_{1}-n_{1}\theta_{2})\wedge\partial^{-1}(m_{1}\theta_{4}-m_{2}\theta_{3})\nonumber\\
&&+(m_{2}\theta_{3}-m_{1}\theta_{4})\wedge\partial^{-1}(n_{2}\theta_{1}-n_{1}\theta_{2}))
\wedge(2{\cal{A}}Q+{\cal{A}}R_{x})\nonumber\\
&&+((n_{2}\theta_{1}-n_{1}\theta_{2}-m_{2}\theta_{3}+m_{1}\theta_{4})\wedge\partial^{-1}Q\nonumber\\
&&+\partial^{-1}(n_{2}\theta_{1}-n_{1}\theta_{2}-m_{2}\theta_{3}+m_{1}\theta_{4})\wedge Q)\wedge(2{\cal{A}}R+{\cal{A}}Q_{x})dx\nonumber\\
&=&\int (f\wedge \partial^{-1}2g+\partial^{-1}f\wedge 2g)\wedge (2{\cal{A}}Q+{\cal{A}}R_{x})\nonumber\\
&&+2(f+g)\wedge({\cal{A}}R_{x}+2{\cal{A}}Q)\wedge({\cal{A}}Q_{x}+2{\cal{A}}R)\nonumber\\
&&+((f+g)\wedge\partial^{-1}Q+\partial^{-1}(f+g)\wedge Q)\wedge(2{\cal{A}}R+{\cal{A}}Q_{x})dx\nonumber\\
&=&\int \partial^{-1}(2g+R)\wedge(Q\wedge({\cal{A}}Q_{x}+2{\cal{A}}R)-R\wedge({\cal{A}}R_{x}+2{\cal{A}}Q))\nonumber\\
&&+(2g+R)\wedge(-\partial^{-1}R\wedge({\cal{A}}R_{x}+2{\cal{A}}Q)+\partial^{-1}Q\wedge({\cal{A}}Q_{x}+2{\cal{A}}R)\nonumber\\
&&+2({\cal{A}}R_{x}+2{\cal{A}}Q)({\cal{A}}Q_{x}+2{\cal{A}}R))dx\nonumber\\
&=&\int \partial^{-1}(2g+R)\wedge(\partial^{-1}R\wedge({\cal{A}}R_{xx}+2{\cal{A}}Q_{x})-\partial^{-1}Q\wedge({\cal{A}}Q_{xx}+2{\cal{A}}R_{x})\nonumber\\
&&-2({\cal{A}}R_{xx}+2{\cal{A}}Q_{x})({\cal{A}}Q_{x}+2{\cal{A}}R)-2({\cal{A}}R_{x}+2{\cal{A}}Q)({\cal{A}}Q_{xx}+2{\cal{A}}R_{x}))dx\nonumber\\
&=&\int\partial^{-1}(2g+R)(\partial^{-1}R\wedge(2{\cal{A}}Q_{x}+4{\cal{A}}R)-\partial^{-1}Q\wedge(2{\cal{A}}R_{x}+4{\cal{A}}Q)\nonumber\\
&&-2\partial^{-1}R\wedge({\cal{A}}Q_{x}+2{\cal{A}}R)-2({\cal{A}}R_{x}+2{\cal{A}}Q)\wedge \partial^{-1}Q)dx.\nonumber\\
&=&0,\nonumber
\end{eqnarray*}
where we use $f-g=R$ for short. Thus, ${\cal L}$ given by \eqref{?} is a Hamiltonian operator.

Finally  we prove the compatibility of ${\cal K}$ and ${\cal L}$, which is equivalent to
\begin{eqnarray}
\mbox{pr}\, v_{{\cal L}\theta}(\Theta_{\cal K})+\mbox{pr}\, v_{\cal K \theta}(\Theta_{\cal L})=\mbox{pr}\,v_{\cal K \theta}(\Theta_{\cal L})=\mbox{pr}\,v_{\cal K \theta}(\Theta_{\cal J})+\mbox{pr}\,v_{\cal K \theta}(\Theta_{\cal F})=0.
\end{eqnarray}
To this end, we notice
\begin{eqnarray*}
-\mbox{pr}\,v_{\cal K \theta}(\Theta_{\cal J})&
=&\int (2(\theta_{4}\wedge\theta_{2})_{x}+(\theta_{1}\wedge\theta_{3})_{x}+\theta_{1}\wedge\theta_{2}+\theta_{3}\wedge\theta_{4})\wedge \partial^{-1}(n_{2}\theta_{4}-m_{2}\theta_{2})\nonumber\\
&&+((\theta_{4}\wedge\theta_{2})_{x}+2(\theta_{1}\wedge\theta_{3})_{x}-\theta_{1}\wedge\theta_{2}-\theta_{3}\wedge\theta_{4})\wedge\partial^{-1}(m_{1}\theta_{1}-n_{1}\theta_{3})\nonumber\\
&&+((\theta_{1}\wedge\theta_{2})_{x}-\theta_{1}\wedge\theta_{3}-\theta_{2}\wedge\theta_{4})\wedge \partial^{-1}(m_{1}\theta_{4}-m_{2}\theta_{3})\nonumber\\
&&+((\theta_{4}\wedge\theta_{3})_{x}+\theta_{1}\wedge\theta_{3}+\theta_{2}\wedge\theta_{4})\wedge \partial^{-1}(n_{2}\theta_{1}-n_{1}\theta_{2}) dx\nonumber\\
&=&\int (-(\theta_{1}\wedge \theta_{2}+\theta_{3}\wedge \theta_{4})\wedge \partial^{-1} Q +(\theta_{1}\wedge\theta_{3}+\theta_{2}\wedge\theta_{4})\wedge \partial^{-1}R\nonumber\\
&&\-(2\theta_{4}\wedge\theta_{2}+\theta_{1}\wedge\theta_{3})\wedge(n_{2}\theta_{4}-m_{2}\theta_{2})-(\theta_{1}\wedge\theta_{2})\wedge(m_{1}\theta_{4}
-m_{2}\theta_{3})\nonumber\\
&&-(\theta_{4}\wedge\theta_{2}+2\theta_{1}\wedge\theta_{3})\wedge(m_{1}\theta_{1}-n_{1}\theta_{3})+(\theta_{3}\wedge\theta_{4})\wedge(n_{2}\theta_{1}-n_{1}\theta_{2})dx\nonumber\\
&=&\int -(\theta_{1}\wedge \theta_{2}+\theta_{3}\wedge \theta_{4})\wedge \partial^{-1} Q+(\theta_{1}\wedge\theta_{3}+\theta_{2}\wedge\theta_{4})\wedge\partial^{-1} R dx.\nonumber
\end{eqnarray*}
and
\begin{eqnarray*}
 -\mbox{pr}\,v_{\cal K \theta}(\Theta_{\cal F})&=&\int (\theta_{1}\wedge(\theta_{3x}-\theta_{2})+\theta_{2}\wedge (\theta_{1}+\theta_{4x})-\theta_{3}\wedge (\theta_{1x}+\theta_{4})-\theta_{4}\wedge (\theta_{2x}-\theta_{3}))\nonumber\\
&&\wedge (2{\cal{A}} Q+\partial{\cal{A}}R)+({\cal{A}}Q_{x}+2{\cal{A}}R)\wedge((\theta_{2x}-\theta_{3})\wedge\theta_{1}-(\theta_{1x}+\theta_{4})\wedge\theta_{2}\nonumber\\
&&+(\theta_{1}+\theta_{4x})\wedge\theta_{3}-(\theta_{3x}-\theta_{2})\wedge\theta_{4})dx.\nonumber\\
&=&\int ((\partial^{2}-4)(\theta_{1}\wedge \theta_{2}+\theta_{3}\wedge \theta_{4}))\wedge {\cal{A}}Q -((\partial^{2}-4)(\theta_{1}\wedge\theta_{3}+\theta_{2}\wedge\theta_{4}))\wedge {\cal{A}}R dx \nonumber\\
&=&\int (\theta_{1}\wedge \theta_{2}+\theta_{3}\wedge \theta_{4})\wedge \partial^{-1} Q-(\theta_{1}\wedge\theta_{3}+\theta_{2}\wedge\theta_{4})\wedge\partial^{-1} R dx.\nonumber
\end{eqnarray*}
Therefore we arrive at
\begin{eqnarray}
\mbox{pr}\,v_{\cal K \theta}(\Theta_{\cal L})=\mbox{pr}\,v_{\cal K \theta}(\Theta_{\cal J}+\Theta_{\cal F})=0,\nonumber
\end{eqnarray} so ${\cal K}$ and $\cal{L}$ are two compatible Hamiltonian operators.

\bigskip
\noindent
{\bf Acknowledgments}

We would like to thank Decio Levi for interesting discussions. NHL and QPL are supported by the National
Natural Science Foundation of China (grant numbers: 10971222 and 11271366) and the Fundamental Research Funds for Central Universities. ZP  thanks the  China University of Mining and Technology for its support during his visit to Beijing where this work has been carried out.


\bigskip


\begin{thebibliography}{20}
\bibitem{Holm}R. Camassa and D. D. Holm, An integrable shallow water equation with peaked solitons, Phys. Rev. Lett. {\bf 71} (1993) 1661-1664.
\bibitem{Hyman}R. Camassa, D. D. Holm and J. M. Hyman, A new integrable shallow water equation, Adv. Appl. Mech. {\bf 31} (1994) 1-33.
\bibitem{Degasperis}A. Degasperis and M. Procesi, Asymptotic integrability, in: Symmetry and Perturbation Theory��  A. Degasperis and G. Gaeta (eds.),  Singapore: World Scientific (1999), pp 23-37.
\bibitem{gas}A. Degasperis, D. D. Holm and A. N. W. Hone, A new integrable equation with peakon solutions, Theor. Math. Phys. {\bf 133} (2002) 1463-74.
\bibitem{for} A. S. Fokas, P. J. Olver and P. Rosenau, A plethora of integrable bi-Hamiltonian equations, in: Algebraic Aspects of Integrable Systems: In Memory of Irene Dorfman, A. S. Fokas and I. M. Gelfand (eds.), Birkhauser: Boston (1996), pp93-101.
\bibitem{fokas} A. S. Fokas, On a class of physically important integrable equations,  Physica D {\bf 87} (1995) 145-150.
\bibitem{fuch} B. Fuchssteiner, Some tricks from the symmetry-toolbox for nonlinear equations: generalizations of the Camassa-Holm equation, Phys. D {\bf 95} (1996) 229-243.
\bibitem{Geng} X. G. Geng and B. Xue,  An extension of integrable peakon equations with cubic nonlinearity, Nonlinearity {\bf 22} (2009) 1847-1856.

\bibitem{Geng2} X. G. Geng and B. Xue, A three-component generalization of Camassa-Holm equation with N-peakon solutions, Adv. Math. {\bf 226} (2011) 827-839.
\bibitem{hwang} A. N. W. Hone and J. P. Wang, Prolongation algebras and Hamiltonian operators for peakon equations, Inverse Problems {\bf 19} (2003) 129-145.
\bibitem{Hone} A. N. W. Hone and J. P. Wang, Integrable peakon equations with cubic nonlinearity, J. Phys. A: Math. Theor. {\bf 41} (2008) 372002.

\bibitem{2novikov} N. H. Li and Q. P. Liu, On Bi-Hamiltonian structure of two-component Novikov equation, Phys. Lett. A {\bf 377} (2012) 257-261.

\bibitem{Li} N. H. Li and Q. P. Liu, Bi-Hamiltonian structure of a three-component Camassa-Holm type equation, J. Nonlinear
Math. Phys. {\bf 20} (2013) 126-134.

\bibitem{szm}H. Lundmark and J. Szmigielski, Multi-peakon solutions of the Degasperis-Procesi equation, Inverse problems {\bf 19} (1994) 1241-5.
\bibitem{Novikov} V. S. Novikov, Generalisations of the Camassa-Holm equation, J. Phys. A: Math. Theor. {\bf 42} (2009) 342002 (14 pages).
\bibitem{PJOlver} P. J. Olver, Applications of Lie Groups to Differential Equations, 2nd edn,
 Berlin: Springer (1993).
\bibitem{olver}P. J. Olver  and P. Rosenau,  Tri-Hamiltonian duality between solitons and solitary-waves solutions having compact support, Phys. Rev. E {\bf 53} (1996) 1900-1906.
\bibitem{popowicz} Z. Popowicz, The generalizations of the peakon's systems, talk given at the third international conference on Nonlinear Waves: theory and applications, June 12-15, 2013, Beijing.
\bibitem{Qiao} Z. J. Qiao, A new integrable equation with cuspons and W/M-shape-peaks solitons, J. Math. Phys. {\bf 47} (2006) 112701 (9 pages).
\bibitem{qsy} C. Z. Qu, J. F. Song and R. X. Yao, Multi-component integrable systems and invariant curve flows in certian geometries, SIGMA {\bf 9} (2013) 001 (19 pages).

\bibitem{Qu}J. F. Song, C. Z. Qu and Z. J. Qiao, A new integrable two-component system with cubic nonlinearity, J. Math. Phys. {\bf 52} (2011) 013503 (9 pages).


\bibitem{Tian} K. Tian and Q. P. Liu, Tri-Hamiltonian duality between the Wadati-Konno-Ichikawa hierarchy and Song-Qu-Qiao hierarchy, J. Math. Phys. {\bf 54} (2013) 043513 (10 pages).



\bibitem{xia} B. Q. Xia and Z. J. Qiao, Integrable multi-component Camassa-Holm system, {\textsc{arXiv: 1310.0268v1}}.


\end{thebibliography}
\end{document}